\documentclass{llncs}

\usepackage{breakurl} 
\usepackage{makeidx}
\usepackage{graphicx,latexsym,alltt}
\usepackage[final]{hyperref}
\usepackage{amssymb,amsmath}
\usepackage{algorithm,algorithmic}
\usepackage{color}
\usepackage{url}
\usepackage{listings}
\usepackage{multirow} 
\usepackage{enumitem} 

\usepackage{multicol}
\usepackage{lipsum}

\usepackage{subcaption}

\usepackage{wrapfig}

\newcommand{\biblio}[1]{../../../../../../Biblio/#1}

\definecolor{blank}{rgb}{0.55,0.55,0.55}

\long\def\comment#1{}

\renewcommand{\phi}{\varphi}

\def\defemb#1#2{\expandafter\def\csname #1\endcsname
                              {\relax\ifmmode #2\else\hbox{$#2$}\fi}}
\defemb{cA}{{\cal A}}
\defemb{cB}{{\cal B}}
\defemb{cC}{{\cal C}}
\defemb{cD}{{\cal D}}
\defemb{cE}{{\cal E}}
\defemb{cF}{{\cal F}}
\defemb{cG}{{\cal G}}
\defemb{cH}{{\cal H}}
\defemb{cI}{{\cal I}}
\defemb{cJ}{{\cal J}}
\defemb{cL}{{\cal L}}
\defemb{cM}{{\cal M}}
\defemb{cN}{{\cal N}}
\defemb{cO}{{\cal O}}
\defemb{cP}{{\cal P}}
\defemb{cR}{{\cal R}}
\defemb{cS}{{\cal S}}
\defemb{cT}{{\cal T}}
\defemb{cU}{{\cal U}}
\defemb{cV}{{\cal V}}
\defemb{cW}{{\cal W}}
\defemb{cX}{{\cal X}}
\defemb{cZ}{{\cal Z}}

\makeatletter
\newenvironment{prog}{\vspace{1.0ex}\par
\obeylines\@vobeyspaces\tt}{\vspace{1.0ex}\noindent
}
\makeatother
\newcommand{\startprog}{\begin{prog}}
\newcommand{\stopprog}{\end{prog}\noindent}

\catcode`\_=\active
\let_=\sb
\catcode`\_=12
\makeatother

\newif\ifpaperVersion
\paperVersionfalse
\ifpaperVersion
	\newcommand{\ignore}[1]{}
	\newcommand{\deleted}[1]{}
	
	\newcommand{\pending}[1]{}
	\newcommand{\done}[1]{}
	\newcommand{\doubt}[1]{}
	\newcommand{\josep}[1]{}
	\newcommand{\david}[1]{}
	\newcommand{\sergio}[1]{}
	\newcommand{\tama}[1]{}
\else
	\definecolor{ignoreColor}{rgb}{1,0.5,0}
	\definecolor{pendingColor}{rgb}{0.2,0.7,0.2}
	\definecolor{doneColor}{rgb}{0.7,0.2,0.7}
	\definecolor{doubtColor}{rgb}{0.6,0.6,0.4}
	\definecolor{josepColor}{rgb}{0.2,0.6,0.6}
	\definecolor{davidColor}{rgb}{0.6,0.2,0.6}
	\definecolor{sergioColor}{rgb}{1.0,0.5,0.0}
 	\definecolor{tamaColor}{rgb}{0.2,0.8,1.0}

	\newcommand{\ignore}[1]{\textcolor{ignoreColor}{\#Ignored: #1}}
	\newcommand{\deleted}[1]{\textcolor{red}{\#Deleted: #1}}
	
	\newcommand{\pending}[1]{\textcolor{pendingColor}{\#Pending: #1}}
	\newcommand{\done}[1]{\textcolor{doneColor}{\#Done: #1}}
	\newcommand{\doubt}[1]{\textcolor{doubtColor}{\#Doubt: #1}}
	\newcommand{\josep}[1]{\textcolor{josepColor}{\#JJJ: #1}}
	\newcommand{\david}[1]{\textcolor{davidColor}{\#DDD: #1}}
	\newcommand{\sergio}[1]{\textcolor{sergioColor}{\#SSS: #1}}
	\newcommand{\tama}[1]{\textcolor{tamaColor}{\#TTT: #1}}
\fi

\usepackage{listings}
\usepackage[utf8]{inputenc}
\usepackage{courier}
\usepackage[T1]{fontenc}
\usepackage[scaled]{couriers}

\usepackage[utf8]{inputenc}
\usepackage[T1]{fontenc}
\usepackage[svgnames]{xcolor}
\usepackage[scaled=0.84]{beramono}
\usepackage{listings}

\makeatletter
\newcommand{\MyChange}[2]{%
  \extractcolorspec{.}\MyChange@CurrentColor
  \extractcolorspec{#2}\MyChange@TestColor
  \ifx\MyChange@CurrentColor\MyChange@TestColor
    {\bfseries\color{green!33!black} #1 }%
  \else
    {#1 }%
  \fi%
}
\makeatother

\lstdefinelanguage{Erlang}
{
classoffset=1,
keywords={},
  keywords = [2]{receive, if, case, try, module, export, all, spawn, end, fun, when, after, spec, of,  compile, true, false, assertEqual, assertError, include, lib},
  keywordstyle=\textbf,
  keywordstyle=[2]\color{blue},
classoffset=2,
  otherkeywords={
    ;, ",\,,\#, ., (, ), \{, \}, [,],|,||,+,-,*,/, <,>, >=, =<,->, =, ==, =:=, _,?
  },
  classoffset=0,
  sensitive=true,
   morecomment=[l]{\%},
     commentstyle=\color{red}\emph,
  stringstyle=\color{black},
  morestring=[b]',
}

\lstdefinelanguage{none}
{
keywords={},
keywords = [2]{},
otherkeywords={},
morekeywords={}
  ,classoffset=1
  ,sensitive=true
  ,stringstyle=\color{black}
  ,morestring=[b]',
  numbers=none,
  ,literate=
}

\begin{document}

\frontmatter
\pagestyle{headings}
\addtocmark{WFLP}

\pagenumbering{arabic}

\title
{
Enhancing POI testing approach through\\ the use of additional information
 \thanks
 {
 This work has been partially supported by MINECO/AEI/FEDER (EU)
under grant TIN2016-76843-C4-1-R,
and by the \emph{Generalitat Valenciana} under grant
PROMETEO-II/2015/013 (SmartLogic).
Salvador Tamarit was partially supported by the \emph{Conselleria de
Educaci\'on, Investigaci\'on, Cultura y Deporte de la Generalitat
Valenciana} under grant APOSTD/2016/036.
  }
}
\titlerunning{Enhancing POI testing approach through the use of additional information}

\author{Sergio P\'erez \and Salvador Tamarit}
\institute
{
Departament de Sistemes Inform\`atics i Computaci\'o\\
Universitat Polit\`ecnica de Val\`encia\\
Cam\'i de Vera s/n\\
E-46022 Val\`encia, Spain\\
 \email{\{serperu,stamarit\}@dsic.upv.es}
}

\maketitle


\begin{abstract}
Recently, a new approach to perform regression testing has been defined: the point of interest (POI) testing. A POI, in this context, is any expression of a program. The approach receives as input a set of relations between POIs from a version of a program and POIs from another version, and also a sequence of input functions, i.e. test cases. Then, a program instrumentation, an input test case generation and different comparison functions are used to obtain the final report which indicates whether the alternative version of the program behaves as expected, e.g. it produces the same values or it uses less CPU/memory. In this paper, we explain how we can improve the POI testing approach through the use of common stack traces and a more sophisticated tracing for calls. These enhancements of the approach allow users to identify errors earlier and easier. Additionally, they enable new comparison modes and new categories of reported unexpected behaviours. 
\end{abstract}



\setlength{\columnsep}{1cm}

\begin{lstlisting}[float, language=none]
\end{lstlisting}

\section{Introduction}
\label{sec:intro}


During its useful lifetime, a program might evolve many times. Each evolution is often composed of several changes that produce a new release of the software. There are multiple ways to control that these changes do not modify the behaviour of any part of the program that was already correct. Most of the companies rely on \emph{regression testing} \cite{YooH12} to assure that a desired behaviour of the original program is kept in the new version, but there exist other alternatives such as the static inference of the impact of changes \cite{jumpertz2010using}.

Even when a program is perfectly working and it fulfils all its functional requirements, sometimes we still need to improve parts of it. There are several reasons why a released program needs to be modified. For instance, improving the maintainability or efficiency; or for other reasons such as obfuscation, security improvement, parallelization, distribution, platform changes, and hardware changes, among others. 
Although regression testing should be ideally done after each change, in real projects the methodology is less "real". As reported in \cite{EngR10}, only 10\% of the companies do regression testing daily. This means that, when an \emph{unexpected behaviour} (UB) is detected, it can be hidden after a large number of subsequent changes. The authors also claim that this long-term regression testing is mainly due to the lack of time and resources.

Programmers that want to check whether the semantics of the original program remains unchanged in the new version usually create a test suite. 
There are several tools that can help in all this process. For instance, Travis CI can be easily integrated in a GitHub repository so that each time a pull request is performed, the test suite is launched.
\emph{Point of interest} (POI) testing \cite{insa2017LOPSTR,insa2017LOPSTR_LNCS,insa2018_arxiv} (briefly described in section \ref{sec:poi_testing}) is an alternative and complementary approach that creates an automatic test suite to do regression testing: 
(i) An alternative approach because it can work as a standalone way without the need of using other techniques. 
Therefore, POI testing can check the evolution of the code even if no test suite has been defined. (ii) A complementary approach because it can also be used to complement other techniques, e.g. unit testing, providing a major reliability in the assurance of behaviour preservation.

In the context of debugging, programmers often use breakpoints to observe the values of an expression during an execution. POI testing makes this feature available in testing, allowing users to easily focus the test cases on one or more specific points without modifying the source code (as it happens when using asserts) or adding more code (as it happens in unit testing).
POI testing introduces the ability to specify kind-of breakpoints in the context of testing. A POI can be any expression in the code (e.g., a function call) meaning that we want to check the behaviour of that expression, e.g. the POI \texttt{\{module.erl, 5, \{var, 'A'\}, 2\}\}} refers to the second occurrence\footnote{When this argument is omitted, it is assumed to be the first occurrence of the expression.} of variable \texttt{A} in the fifth line of file \texttt{module.erl}. 
Although they handle similar concepts, POIs are not exactly like breakpoints, since their purpose is different. Breakpoints are used to indicate where the computation should stop, so users can inspect variable values or control statements. In contrast, a POI defines an expression whose sequence of evaluations to values must be recorded, so that users can check the expected behaviour (by value comparison) after the execution. 


Although POI testing is a complete approach that allows users to control how, when and what should be compared, there are still some improvements which can increase its usability. In this paper, we present two enhancements of the approach that enrich it with new information, new features and new comparison modes. These enhancements share a common framework, described in Section \ref{sec:general}, which can be used to define new and more complex improvements of the technique. In particular, the enhancements presented in this paper improve the POI testing approach by the use of better call traces, described in Section \ref{sec:call}, and stack traces, described in Section \ref{sec:stack}. The goal of both enhancements is to provide users with more information and to ease the discovery of UB sources. In this paper, we provide concrete examples by using Erlang as the running language.

On the one hand, the call trace enhancement provides more information about calls in such a way that, when an UB is detected in a call expression, users can automatically compare the call arguments that produced this UB. Therefore, users can know earlier if the cause of the UB is a discrepancy between these arguments, or a behaviour change in the called function. 
On the other hand, we define a enhancement which makes use of the common stack traces, i.e. the ones showed in error report. Stack trace are a common element used while debugging buggy code. Therefore, it seems natural to incorporate them into the POI testing approach. 
Stack traces can be compared to find earlier whether an UB comes from some differences in the followed execution paths. 
Using this discrepancy it is easier to point to the code which causes the UB and start a debugging process from there. 

\section{POI testing}
\label{sec:poi_testing}

%

In POI testing, (i) the programmer identifies a POI and a set of \emph{input functions} whose invocations should evaluate the POI. 
Then, by using some automatic test case generation technique, (ii) the approach automatically generates a test suite that tries to cover all possible paths that reach the POI (trying also to produce execution paths that evaluate the POI several times). 
Therefore, in POI testing, the \emph{input of a test case} (ITC) is defined as a call to an input function with some specific arguments, and the output is the sequence of those values the POIs are evaluated to during the execution of the ITC. For the sake of disambiguation, in the rest of the paper we use the term \emph{traces} to refer to these sequences of values. 
Next, (iii) the test suite is used to automatically check whether the behaviour of the program remains unchanged across new versions
\footnote{Steps (ii) and (iii) could also be executed in parallel. Here, for the sake of simplicity, we only consider the sequential execution of these steps.}. 
Finally, (iv) the user is provided with a report about the success or failure of these test cases. 

Note that the technique allows the definition of multiple POIs \cite{IPST18}.
With this feature, 
users can trace several (and maybe unrelated) functionalities in a single run. 
Additionally, users can strengthen the quality of their test suite by checking behaviour preservation in more than one point. 
Finally, this feature is needed in those cases where a POI in one version is associated with more than one POI in another version (e.g., when a POI in the final source code is associated with two or more POIs in the initial source code due to a refactoring or a removal of duplicated code). 

An example of a POI tester is the tool named \texttt{SecEr} (\emph{Software Evolution Control for Erlang}), which is publicly available at: \url{https://github.com/mistupv/secer}. 
All the analyses performed by \texttt{SecEr} are transparent to the user. The only task in \texttt{SecEr} that requires user intervention is identifying suitable POIs in both the old and the new versions of the program. \texttt{SecEr}  allows to define test configuration files to ease all this process and also to make it reusable.
The interested readers are referred to \cite{IPST18} where they can found an extensive discussion about the similarities with similar tools and how to deal with concurrency.

In the following, we introduce all the concrete details of the approach needed to understand the enhancements presented in this paper. It has been divided in three items, the inputs that the approach needs, a summarized description of its internals, and the outputs it produces.

\begin{itemize}

\item \textbf{Inputs:} POI testing approach needs two parameters at least to be able to operate. Additionally, the POI testing approach can also be run with some specific comparison and report functions. The comparison and report functions are explained within the internals and the outputs of the approach, respectively. 

\begin{itemize}

\item \emph{POI relation.} It relates POIs from two different versions of the same program. In Erlang, it is represented by a list of tuples. Each one of these tuples contain two POIs, one of each version of the program. For instance, the POI relation \texttt{[\{\{old.erl, 14,  \{var, 'X'\}\}, \{new.erl, 14, \{var, 'BetterName'\}, 1\}\}, \{\{old.erl, 130, case\}, \{new.erl, 145, if\}\}]} defines a relation between two POIs. The first one is indicating that the variable \texttt{X} at line 14 has been renamed to \texttt{BetterName}. The second one indicates that a \texttt{case} expression has been changed by an \texttt{if} expression and their lines has also changed from 130 to 145. 

\item \emph{Input functions.} They are the entry points of the test cases. In Erlang, it is a list of function names. For examples, the list \texttt{[main/2, test_1/0]} defines two entry points for the approach, i.e. all the ITCs generated for the approach are calls to one of these functions. In concrete, function \texttt{main/2} requires the generation of concrete arguments which are not provided by the user. This is further discussed in the  internals of the approach. 
 On the other hand, function \texttt{test_1/0} has no arguments, so the approach simply run it once, and no more ITCs can be generated for it. Functions like this one are usually unit tests. Therefore, unit tests cannot be used in the POI testing approach to perform regression testing, instead they can only be used to debug failing test cases. 

\end{itemize}

\item \textbf{Internals:} There are three main stages of the approach. First, how the traces are built, then how they are compared and finally how new ITCs can be generated.

\begin{itemize}

\item \emph{Trace building.} The basis of the POI testing approach is the tracing of some POIs during the evaluation of a concrete ITC. For this reason, it is needed a way to both, create and collect these traces. The first one is usually done by a program instrumentation, like the one defined in \cite{IPST18} for Erlang. This program instrumentation builds tuples of the form $(POI, value)$\footnote{Note that $value$ is not necessarily the value the expression is evaluated to. It can also be other type of values, e.g. the time needed to evaluate it.} which are produced during the evaluation of certain ITC. However, it is not enough with this, and we need a way to collect and sort these traces. In Erlang, a trace server is used in the following way: all the tuples produced by the instrumentation are sent to a server which collects and sorts them building the final trace of a concrete ITC evaluation. 

\item \emph{Trace comparison.} Once the traces are generated and stored, the next step is to compare them in order to infer if there is any UB\footnote{The observed UBs are represented and identified using literals, e.g. the atom \texttt{slower} could be used to represent an UB that occurs when an expression took more time to evaluate in the new version than in the old one. The UB representations are defined during the comparison process as it is then when the UBs are found.}. In case some UB is found, its UB type together with an UB report is generated. These are several ways of comparing the traces. The most relevant techniques to compare multi-POI traces are described in \cite{IPST18}. The main distinction between them is whether the traces are compared as a whole or independently for each POI. For example, consider two versions of a program, with two POIs each one, i.e. first version (\texttt{POI_1} and \texttt{POI_2}) and second version (\texttt{POI_1'} and \texttt{POI_2'}). These POIs are related in the following way: \texttt{POI_1} is related with \texttt{POI_1'} and \texttt{POI_2} is related with \texttt{POI_2'}. After the execution of a concrete ITC in bot version we obtain the following traces: \texttt{[\{POI_1, 3\}, \{POI_2, 4\}]} and \texttt{[\{POI_2', 4\}, \{POI_1', 2\}]}. If they are compared as a whole, the UB reported is that a trace from POI \texttt{POI_1} was expected but a trace from POI \texttt{POI_2} was generated. On the other hand, when they are compared independently, independent traces are generated for each POI. Therefore, for this example, there is not any UB when comparing the traces of \texttt{POI_2}, but there is an UB in \texttt{POI_1}'s traces, i.e. value 3 was expected but value 2 was generated. There is an optional input parameter that allows users to define their own comparison functions. A user-define comparison function should receive two traces as input, i.e. the traces to be compared, and should return either $true$ or a tuple of the form $(UB\_Type, UB\_Report)$, where $UB\_Type$ is the type of the observed UB and $UB\_Report$ is a string to be printed when the UB is reported to the user.

\item \emph{ITC generation.} The POI testing approach starts by generating an ITC for each input function. However, in order to reinforce the obtained UB report, each time an ITC is compared new ITCs for the input functions should be generated according to the comparison result. These new ITCs are usually based on the results of the previously compared ITCs. In this way, when an UB is found for a concrete ITC, we can generate new ITCs based on this ITC so they are more propitious to generate the same or other UBs. In \cite{IPST18}, an ITC generation based on the mutation of the arguments of the ITC is described. Alternative generators can be used, but they should ideally take into account the result of the trace comparison in order to obtain better results for the POI testing approach.

\end{itemize}

\item \textbf{Outputs:} The POI testing approach produces as output a collection of ITCs together with the result of the trace comparison. When none of the ITC evaluated have generated an UB, users are informed of the successful result by adding also some additional information like the number of ITCs evaluated. Conversely, when one or more UBs have been observed, users get a report that can be configured in several ways. For example, the UB report can show only an ITC sample for each observed UB type in two ways: without distinguish between POIs or for each individual POI. Another alternative is to show all the  ITCs where an UB has been observed. 
Additionally, users can define report functions to choose the reported information. 
A report function receives the comparison result and defines what information should appear in the UB report. 
Finally, some or all the failing ITCs can be stored somewhere so they can be reused after UB debugging to check whether the observed UBs have been mended.

\end{itemize}

\section{Enhancing POI testing with additional information }
\label{sec:general}


This section introduces a general overview of the enhancements that we have defined for the POI testing approach. Therefore, Sections \ref{sec:call} and \ref{sec:stack} are special cases of the general methodology explained here. 


\subsection{Obtaining and merging additional information with current traces}
\label{sec:gen-trace}


POI testing uses POI traces to check whether some UBs exist across several program versions. We represent each element of this trace as a tuple: $(POI, value)$
In this paper, we propose an extension where some additional information is attached to each \emph{trace element} (TE). Therefore, we need to extend this  representation in order to be able to refer to this additional information. Moreover, we want to do this extension in a way that future trace enhancements follow the same scheme. For this reason, we have chosen a mapping function, represented as $ai$, e.g. $ai(st)$ will return the stored information about stack traces. Thus, in the enhanced approach a TE is a triplet $(POI, value, ai)$. 

Therefore, a concrete implementation of an enhanced POI testing approach should be able to produce traces which TEs are not tuples but these triplets. As mentioned in Section \ref{sec:poi_testing}, the TEs, i.e. the tuples, are sent when the instrumented code is executed. Then, these TEs are collected and stored by the tracer which produces the final trace.  
Thus, the task of each concrete enhancement of the POI testing approach is to build these triplets storing in $ai$ all the particular information needed for the posterior trace comparison, e.g. stack traces. 
In other words, for each enhancement we need to build a specific code instrumentation and a tracer. 

\subsection{Using enhanced traces to check program behaviour}
\label{sec:gen-use}

\begin{wrapfigure}[13]{r}{0.62\textwidth}
\vspace{-30pt}
\begin{center}
\begin{lstlisting}[basicstyle=\ttfamily\scriptsize, frame=single, 
%caption={Comparison function which ignores the callee}, label=lst:fc_ign_callee, 
language=erlang,
numbers=left, stepnumber=1,escapechar=@]
cf_general(TO, TN, VEF, TECF, UBRM) -> 
  cf_general(TO, TN, VEF, TECF, UBRM, []).

cf_general([], [], _, _, _, _) -> 
  true;
cf_general([TOE | TO], [TNE | TN], VEF, TECF, UBRM, His) ->
  case TECF(VEF, TOE, TNE) of 
    true ->  
      cf_general(TO, TN, VEF, TECF, UBRM, [{TOE ,TNE} | His]);
    UBT -> 
      {
        UBT, 
        (dict:fetch(UBT, UBRM))(TOE, TNE, lists:reverse(His))@\label{lst:gen_cf_1}@
      } 
  end.
\end{lstlisting}
\vspace{-20pt}
\end{center}
\caption{General comparison function}
\label{fig:gen_cf}
\vspace{-10pt}
\end{wrapfigure}
POI testing allows using any comparison function. This feature gives users a complete freedom to configure the testing and/or debugging process in the best way according to their needs. However, it is common that users always rely on the same type of comparison functions. Therefore, an implementation of the POI testing approach usually provides some default comparison functions. 
Then, in order to generically deal with all possible enhancements of the POI testing approach, we define a default comparison function.

The general comparison function for our running language Erlang is depicted in Figure \ref{fig:gen_cf}.  The first two parameters are the usual ones, i.e. the whole traces of two program version, \texttt{TO} and \texttt{TN}. The rest of parameters of this function are introduced below.

\begin{itemize}

\item \textbf{Value-extractor function \texttt{(VEF)}}: This function extracts the concrete value that will be used to compare each TE. It takes a TE as input an returns the comparison values. For example, function \texttt{fun(TE) -> TE end} uses the whole TE to check UBs, i.e. its POI, its value and all its additional information. On the other hand,  function \texttt{fun(\{POI, V, AI\}) -> \{POI, V, dict:fetch(st, AI)\} end}\footnote{Function \texttt{dict:fetch(Key, Dict)} returns the value associated with \texttt{Key} in dictionary \texttt{Dict}.} uses only the stack trace information and ignores the rest of additional information. The default value for \texttt{VEF} is \texttt{fun(\{\_, V, AI\}) -> \{V, AI\} end}\footnote{We assume that the POI comparison is previously performed by a POI-relation check function.}.

\item \textbf{Trace-element comparison function (\texttt{TECF})}: In order to allow users to check UBs in different ways and not only a plain equality function, i.e. operator \texttt{==}, we add a comparison function for each pair of TEs. This function takes as input the \texttt{VEF} and the TEs of both versions of the program. Internally, it applies \texttt{VEF} to each TE and decides whether the observed behaviour is the expected or not. In case of an UB is found, its type is returned. For example, \texttt{fun(VEF, TOE, TNE) -> case compare(VEF(TOE), VEF(TNE)) of gt -> true; eq -> same; lt ->  downgrade end end}, is a simple example where function \texttt{compare/2} is used to check whether a reduction in some performance indicator is obtained. Then, when it is not obtained, either a \texttt{same} or a \texttt{downgrade} UB type is returned. The default value for \texttt{TECF} is \texttt{fun(VEF, TOE, TNE) -> VEF(TOE) == VEF(TNE) end}.
  
\item \textbf{{Unexpected-behaviour report mapping (\texttt{UBRM})}}: When an UB is detected, a specific report should be generated, i.e. a message should be provided to users. Therefore, this parameter allows users to specify how the POI tester should react to a particular UB. In this case, we use a mapping, because it allows easy redefinitions and additions of UBs reports. The mapping returns a function for a given \emph{unexpected behaviour type} \texttt{UBT}. The returned function builds a string using the TEs of both versions of the program and the previous compared TEs, i.e. \texttt{His}. For example, expression \texttt{dict:fetch(diff_values, UBRM)} in line \ref{lst:gen_cf_1} of Figure \ref{fig:gen_cf}, can return a function like \texttt{fun(\{P, V1, _\}, \{P, V2, _\}, His) -> "Value for P differ. V1 vs V2. Previous trace elements: His" end}\footnote{This is not a valid Erlang string. We do not show here the actual implementation for the sake of the presentation simplicity.}. Thus, in this way, users can define the message that should be shown when an UB of type \texttt{diff_values} is found. The default value for \texttt{UBRM} is a mapping that, for any UB type, returns a function that prints all the available information, i.e. the current  and the previous TEs. 

\end{itemize}

There are several modes of using the additional information stored in the TEs and all these modes are defined by the arguments given to the general comparison function (Figure \ref{fig:gen_cf}). Although, given the freedom of the POI testing approach, more specific and user-defined modes are possible, we list the three modes which surely will be more frequently needed by users.

\begin{itemize}

\item \textbf{Additional information is not used during comparison (NUAI).} In this mode, the traced values are the only data used when comparing the TEs. This is the mode that should be used when the additional information is expected to vary due to the differences between program versions or simply due to the type of data it contains. Additionally, this mode can also be used to lighten the comparison process or to reduce the relevance of the additional information. This mode will use a value-extractor function like  \texttt{fun(\{POI, V, _\}) -> \{POI, V\} end} or a particular variant, where the additional information is simply ignored. According on how the additional information is used, we have identified three submodes.
\begin{itemize}

\item \textbf{Additional information is only used to define UB types (NUAI-T).} The additional information is only used to define new types of UBs, but it will not appear in the UB report. This mode is really convenient in such cases where the additional information is too complex or too big, so it will not give a significative feedback to the user. However, when a UB is found it is interesting to consider it to build a specific type of UB, e.g. \texttt{diff_value_same_stack_trace} can characterize those UBs where a difference in the POI values has been found but their stack traces were the same. The trace-element comparison function is the one that should be defined to use this mode. For example, Listing \ref{lst:gen_ub_types} shows a trace-element comparison function which distinguishes between those unexpected values where the stack trace is the same and those where the stack trace is different\footnote{Function \texttt{ai} is defined as \texttt{ai(\{_,_,AI\}) -> AI}.}.  

\begin{lstlisting}[basicstyle=\ttfamily\scriptsize, frame=single, 
caption={Trace-element comparison function which returns different UB types. }, label=lst:gen_ub_types, 
language=erlang,
numbers=left, stepnumber=1,escapechar=@]
fun(VEF, TOE, TNE) -> 
    case VEF(TOE) == VEF(TNE) of 
        true -> true; 
        false -> @\label{lst:lst:gen_ub_types_1}@
            case dict:fetch(st, ai(TOE)) == dict:fetch(st, ai(TNE)) of  @\label{lst:lst:gen_ub_types_2}@
                true -> diff_value_same_stack_trace; @\label{lst:lst:gen_ub_types_3}@
                false -> diff_value_diff_stack_trace @\label{lst:lst:gen_ub_types_4}@
            end
    end 
end
\end{lstlisting}

Categorizing different types of UBs has several benefits in the POI testing approach. First of all, these types can be considered in the ITC generation as a criteria to decide whether an ITC should be mutated or not. In the example above, if we do not distinguish between \texttt{diff_value_same_stack_trace} and \texttt{diff_value_diff_stack_trace}, once a \texttt{diff_value} type has been mutated it can have less chances of being selected to be mutated again. However, with the distinction, each one is treated separately, so they are mutated as separated entities. Additionally, if the final report is enriched with several UB types, users have more feedback that can help while finding the source of the UB.  

\item \textbf{Additional information is only used in the UB report (NUAI-R).} In case we consider that the additional information is not representative enough to categorize new type of UBs, we can use its data only in the reports. This is the less intrusive way of using the additional information, but still a useful way to obtain richer feedback in the final report of each UB. We should add to the UB report mapping a new function associated to each UB type that could benefit from the stored additional information. 
For example, we can store in the UB report mapping a function \texttt{fun(\{P1, V1, AI1\}, \{P2, V2, AI2\}, _) -> "Value for P1 (V1) and for P2 (V2) differ.$\backslash$n Their stacks were:$\backslash$n dict:fetch(st, AI1)$\backslash$n dict:fetch(st, AI2)" end} associated with key \texttt{diff_value}. Therefore, each time there is a difference in the compared values, we also get a feedback on their stack traces although internally it will be treated as a simple \texttt{diff_value} UB type.

\item \textbf{Additional information is used to categorize and report UBs (NUAI-TR).} This submode takes the advantages of both previous submodes. It also involves specific trace-element comparison function and additions in the UB report mapping. As this submode is a conjunction of both previous submodes, it is not further discussed in the particular enhancements presented in Sections \ref{sec:call} and \ref{sec:stack}.

\end{itemize}

\item \textbf{Additional information is used during comparison (UAI).} This mode is the one that gives a major relevance to the additional information. By using this mode, the value and the additional information is compared as a whole. This means that, for instance, even if the compared values are the same, when any pair of elements of the additional information differs, the ITC is reported to be generating an UB. 
This mode is very convenient to early uncover some UBs. It can also be used for performance checking, e.g. the values of the TEs are equal but a performance indicator included in the additional information is revealing some downgrade. 
This mode uses a value-extractor function and a trace-element comparison function which takes into account all or some parts of the additional information.
The amount of information that is finally used to build the UB reports is left to user's choice. 

\item \textbf{The additional information is not attached to the TE, instead considered as an independent trace element (AIT).} 
Finally, in this completely different mode, the additional information is considered as a separated entity and constitutes a single TE as the ones that are generated for the POIs. 
This mode can be similar to the original POI testing approach, however it still needs special instrumentation (to send the new TEs), tracing (to receive and store the new TEs) and maybe some special comparison functions (to take into account their particularities).
This mode is very convenient in such cases where the additional information can be directly used to uncover an UB, avoiding in this way the comparison of several subcomputations.
For instance, if a user places a POI in a call, and the call parameters are compared before comparing the call result, all intermediate TEs are not compared.
This mode can reuse the program instrumentation of modes UAI and NUAI in most of the cases. However, the tracer should be redefined in order to build the new traces accordingly.
Finally,  this mode can be combined in such a way that other additional information is attached to these special TEs forming a hybrid mode which can help the user in some specific scenarios.



\end{itemize}

\section{Enhancement by using improved call tracing}
\label{sec:call}

In this section we explain both, how we have improved the call tracing and how we can incorporate the enhanced call traces to the POI testing approach. 

\subsection{Motivation}
\label{sec:call-mot}


As it is usual in testing and specially in debugging, when an UB is found we still have to find its source to fix it. Unfortunately, the POI testing approach is not an exception. For example, consider a call that is used as POI and the values computed by different version of a program are different. We know that the UB is due to the call, however what we do not know is whether the problem is in the arguments of the call or inside the function called. With the previous versions of POI testing several iterations \footnote{An iteration includes creating and moving POIs and comparing the new traces which need to be recomputed.} would be needed in order to find an answer to this question. With the enhancement that we propose here, we can save some time to users by avoiding these intermediate steps. 
In the new approach, the POIs that are calls will be treated in a special way. This special treatment allows us to directly know where is the source of the UB, i.e. in the argument or in the called function. 
This is just one of the benefits of using an enhanced tracing for calls, but there are more explained in the rest of the section.


\subsection{Improvements in the call tracing}
\label{sec:call-trace}

When we place a POI in a call, we are saying that we are interested in comparing the result of this call, so the standard behaviour of a POI tester is to trace only these values. In this work, we want to create an enhanced trace where not only the result of the call, but also its arguments are traced. Therefore, this enhancement adds to the additional information mapping a new element whose key is \texttt{ca} and whose value is a list that contains the call arguments.


In order to obtain the improved call traces, we have to define a way for sending, receiving and merging the call traces. The main idea is to send the arguments traces before actually performing the call and its result just after. Thus, we should define how the code instrumentation is extended to create this enhanced TEs, and also we should define how the tracer deals with them.


\begin{figure}[t!]
    \centering
    \begin{subfigure}[t]{0.45\textwidth}
  \begin{minipage}{\linewidth}
  \scriptsize{
\scalebox{1.0}{
\setlength\arraycolsep{5pt} 
$\begin{array}{ll}
\mathtt{e(\overline{e_i})} \Rightarrow & \hspace{-6pt}\mathtt{begin}\\
& \mathtt{fv_{ref} = make\_ref(),}\\
& \mathtt{[fv_v | \overline{fv_{v_i}}] ~=~ [e | \overline{e_i}],}\\
& \mathtt{tracer!\{add\_i, POI, fv_{ref}, fv_{v}\},  }\\
& \mathtt{\overline{tracer!\{add\_i, POI, fv_{ref}, fv_{i}\},}  }\\
& \mathtt{fv = fv_v(\overline{fv_{v_i}}),}\\ 
& \mathtt{tracer!\{add, POI, fv_{ref}, fv\},}\\
& \mathtt{fv}\\ 
&  \hspace{-6pt}\mathtt{end}
\end{array}$
}}
  \end{minipage}
\vspace{0.1cm}
\caption{Instrumentation rule for call tracing}
\label{fig:call_rule}
    \end{subfigure}%
    \vline~~~~~~
    \begin{subfigure}[t]{0.55\textwidth}
  \begin{minipage}{\linewidth}
\begin{lstlisting}[basicstyle=\ttfamily\scriptsize,  
numbers=left, stepnumber=1,escapechar=@]
tracer({Stack, Trace}) ->
    receive
        {add_i, POI, Ref, V} ->
            tracer({[{Ref, V} | Stack], Trace});
        {add, POI, Ref, V} ->
            {CalleeArgs, NStack} = 
                remove_same_ref(Ref, Stack),
            tracer({NStack, 
                [{POI, V, store(ca, CalleeArgs)} 
                 | Trace]});
        {add, POI, V} -> @\label{lst:tracer_1}@
            tracer({Stack, [{POI, V} | Trace]}) @\label{lst:tracer_2}@
    end.
\end{lstlisting}
  \end{minipage}
  \vspace{-0.3cm}
\caption{Simplified tracing server}
\label{fig:tracer}
    \end{subfigure}
    \caption{Example of the elements needed to obtain an enhanced call tracer in Erlang}
\end{figure}

The sending process is done thanks to a program instrumentation that enables this double tracing for the call in two steps, i.e. arguments before performing the call and the result just after the call. We show in Figure \ref{fig:call_rule} how this instrumentation can be done in our running language, i.e. Erlang. When the code instrumentation process finds a call, i.e. $\mathtt{e(\overline{e_i})}$, the expression is the replaced by the block expression (\texttt{begin-end}) on the right-hand side. The first expression of this block creates a unique reference (with function \texttt{make_ref/0}) which serves to identify all the traces belonging to the same concrete execution of a call. The result is stored in a free variable, i.e. $\mathtt{fv_{ref}}$\footnote{All free variables used in the rule are represented as $\mathtt{fv_{*}}$. Each one of these free variables is unique in the instrumented module and different to all the original variables of the module.}.  
The second expression evaluates the callee an its arguments\footnote{It could happen that some of these expressions are already POIs or that some of these expressions are calls, e.g. \texttt{f(2, g(1))}. The POI testing approach is ready to handle all these scenarios \cite{IPST18}.} and store its result separately making use of pattern matching facilities, i.e. $\mathtt{[fv_v | \overline{fv_{v_i}}]}$. Then, the next two expressions are sending to the tracer the value of the callee, and each one of its argument. 
The third expression in the block is performing the actual call by using the value of the callee, i.e. $\mathtt{fv_v}$, and the values for the arguments, i.e. $\mathtt{\overline{fv_{v_i}}}$. Then, the result is stored in the fresh variable \texttt{fv}. The fourth expression in the block is sending to the tracer the result of the evaluation, i.e. \texttt{fv}, the POI identifier, i.e. \texttt{POI}, and also the reference that uniquely identifies the call, i.e. $\mathtt{fv_{ref}}$. Note that this reference is the same one that it has been used inside the list comprehension when the values of the callee and the arguments have been sent. However, the atom at the first element of the tuple sent to the tracer is different in each case, i.e. one uses \texttt{add_i} and the other uses \texttt{add}. The reason for this difference is explained latter. Finally, the actual value of the call, i.e. \texttt{fv}, is placed as the last expression to make the whole block evaluate to the expected result.

All the information sent while running the instrumented code is received and merged by the tracer. In Erlang, the tracer is a server which is continuously receiving TEs until the end of the execution or until a timeout is raised. Figure \ref{fig:tracer} shows a simplification of the Erlang function \texttt{tracer/1} which is in charge of this tracing process. The server's state is a tuple containing: 1) a stack, where the callee and arguments are stored in the order they are received, and 2) the trace generated so far. Its body is a receive expression with three clauses: the first one is for the information sent by function calls' callees and arguments, the second one is for the result of the function call, and the third one is for the rest of TEs, i.e. those that do not come from a function call. When a callee or an argument value is received, it is simply stacked. When the call result is received, all its arguments, which are at the top of the stack, are unstacked by using function \texttt{remove_same_ref/2}, and stored\footnote{Function \texttt{store/2} is defined as \texttt{store(Key, Value) -> dict:store(Key, Value, dict:new())}.} in the additional information of the call's TE. Finally, the rest of TEs are simply added to the current trace with an empty additional information mapping.

\subsection{Using the enhanced call tracing to compare traces}
\label{sec:call-use}

Once we have these new traces for the function call, we have to define how we can use them to determine whether the behaviour across different program versions is the expected. In Section \ref{sec:gen-use}, the most useful comparison modes are introduced. Here, we describe the particular requirements needed to implement this enhancement.

\begin{itemize}

\item \textbf{NUAI mode:} The enhanced call trace is compared by only using the call result, i.e. as the non-enhanced POI testing approach does. However, when an UB is found, the callee and the arguments can be used in three ways.

\begin{itemize}

\item \textbf{NUAI-T mode:} In this mode, users can create new UB categories related to the call traces. The trace-element comparison function in Lisinting \ref{lst:gen_ub_types} can be used in this enhancement by doing some small changes. First of all, only the POIs which are calls have the call-arguments additional information. Therefore, when the UB is found (line \ref{lst:lst:gen_ub_types_1}), we should first check that this additional information is available. Then, the access to the mappings in line \ref{lst:lst:gen_ub_types_2} should be change to \texttt{dict:fetch(ca, ai(TOE~|~TNE))}, and the UB types returned in lines \ref{lst:lst:gen_ub_types_3} and \ref{lst:lst:gen_ub_types_4} should be changed to something like \texttt{diff_value_same_call_args} and \texttt{diff_value_diff_call_args} resdpectively.

\item \textbf{NUAI-R mode:} This mode allows users to save the intermediate step mentioned in Section \ref{sec:call-mot}. For instance, instead of just reporting UBs like \emph{"The values of the last trace elements differ"}, it can now produce reports like \emph{"The values of the last trace elements differ and their calls are the same"} and \emph{"The values of the last trace elements differ and their calls are different"}. 


\end{itemize}

\item \textbf{UAI mode:} The enhanced call trace is compared using the call result as well as the callee and the arguments. This means that even when the call results behave as expected, if something in the call arguments is unexpected, an UB is reported. 

\item \textbf{AIT mode:} Once the callee and the arguments values differ, they are directly reported as an UB, without having to compare the rest of the TEs generated between them and the call result. For instance, this can save time and resources when analyzing recursive functions. It is even useful when it has been performed some function/module renaming or some change in the parameter order. For example, by using the value-extractor function \texttt{fun(\{POI, \{callee_args, [_|Args]\}, _\}) -> \{POI, Args\} end}, we are ignoring the callee, so just the arguments are used during comparison.
Note that here we are assuming the existence of \texttt{callee_args} traces, while  they are not introduced in Section \ref{sec:call-trace}. In order to obtain them we should modify Figures \ref{fig:call_rule} and \ref{fig:tracer}. On Figure \ref{fig:call_rule}, the call result is sent without including the unique reference, i.e. it is replaced by $\mathtt{tracer!\{add, POI, fv\}}$. On the other hand, Figure \ref{fig:tracer} needs more modifications. The alternative trace server is depicted in Figure \ref{fig:tracer2}. The server's state is the same but the stack is used here to temporally store all the elements of the call until they are all sent. The first \texttt{receive} clause (lines \ref{lst:tracer2_1}-\ref{lst:tracer2_2}) treats all the information coming from callees and arguments of calls. It distinguishes 2 cases: the top of the stack is an element with the same reference as the received TE, and the rest of cases. Thus, the first one is in charge of storing in the current trace an already-complete call trace, while the second clause is storing a part of a call trace in the stack. The same idea is followed by the other clause (lines  \ref{lst:tracer2_3}-\ref{lst:tracer2_4}).




\end{itemize}


\setlength{\columnsep}{1cm}
\begin{figure}
\hrule
\begin{multicols}{2}
\begin{lstlisting}[tabsize=2,basicstyle=\ttfamily\scriptsize, 
%frame=bt,
%caption={string.erl\\ (optimized version)}, label=lst:string1, 
numbers=left, stepnumber=1,escapechar=@]
tracer({Stack, Trace}) ->
  receive
    {add_i, POI, Ref, V} ->@\label{lst:tracer2_1}@
      case Stack of 
        [{PrevRef, _} | _] when PrevRef /= Ref ->
          {CAs, [PrevPOI]} = 
            remove_same_ref(PrevRef, Stack),
          tracer(
            {[{Ref, V}, POI], 
             [{PrevPOI, {callee_args, CAs}}
              | Trace]});
        _ ->
          tracer(
            {[{Ref, POI, V} | Stack], 
             Trace})
      end;@\label{lst:tracer2_2}@
    {add, POI, V} ->@\label{lst:tracer2_3}@
      case Stack of 
        [{Ref, _} | _] ->
          {CAs, [PrevPOI]} = 
            remove_same_ref(Ref, Stack),
            tracer(
              {[], 
               [{POI, V}, 
                {PrevPOI, {callee_args, CAs}}
                | Trace]});
        [] ->
          tracer({[], [{POI, V} | Trace]})
      end@\label{lst:tracer2_4}@
  end.      
\end{lstlisting}
\end{multicols}
\vspace{0.25cm}
\hrule
\caption{Alternative server that enables independent callee and arguments comparison}
\label{fig:tracer2}
\end{figure}


\section{Enhancement by using stack traces}
\label{sec:stack}

In this section we explain how the POI testing approach is improved by the insertion of stack traces in the POI traces. 

\subsection{Motivation}
\label{sec:stack-mot}

Stack traces are common elements in error reports due to the valuable information they provide to help find the source of a failure. In our context, a failure is not a common bug but an UB. Nevertheless, stack traces can be also really handy in this context. Suppose we are performing POI testing on two programs using a single POI inside a common function that is called from different parts of the program. When the POI tester is run, we can get some reports informing of some UB, for instance, that the POI values are different in a particular execution point. Then, users should start placing POIs in previous stages of the evaluation in order to find the source of the discrepancy. However, it is not clear how to proceed in this debugging process, as there is not enough information about the discrepancy's source. This discrepancy could happen for several reasons. As in Section \ref{sec:call}, one of this reason can be that the calls are not executed with the same arguments. However, if we are using as POI an expression which is not a call, e.g. a case expression, we cannot benefit from the enhanced call trace information. Therefore, we need to explore alternative ways to provide users with some evaluation context when an UB is found. By providing UB reports with stack traces, users can check whether both versions have perform the same calls or not, i.e. they have followed parallel paths. Even when the top of the stack trace is a call with the same arguments, the produced value can differ simply because some of the elements of the rest of the stack trace differ. The discrepancy in the followed paths can be, e.g. due to impure features of the language, like, for instance, the process dictionary in Erlang. Thus, it is not a simple task to identify these discrepancies. With the enhancement proposed here, users can use the stack trace to go directly to the function that start creating the bifurcation on the paths and start a debugging process there. By using special comparison functions, this enhancement can be useful even when some renaming or refactoring process has been performed. 

\subsection{Adding stack traces to the POI traces}
\label{sec:stack-trace}

In a similar way as in Section \ref{sec:call-trace}, we need to modify the standard POI testing approach in order to add stack traces to the POI testing approach. This again involves modifying sending and reception of the TEs. However, in this case, the modifications needed are much simpler. First, the rules used by the instrumentation process need to augment each message send to the tracer with the stack trace, e.g. $\mathtt{tracer!\{add, POI, fv_{ref}, fv\}}$ in Figure \ref{fig:call_rule}. In Erlang, this is done by using \texttt{erlang:get_stacktrace/0}\footnote{Calls to this function are only allowed in the handler of a try-catch expression. We do not considered this particularity for the sake of simplicity of the presentation.}. On the other hand, the reception should be adapted accordingly to process these new TEs. This involves, modifying the receive's clauses. In concrete, we can use a single clause like the one in lines \ref{lst:tracer_1}-\ref{lst:tracer_2} of Figure \ref{fig:tracer}, e.g. \texttt{\{add, POI, V, ST\} -> tracer([\{POI, V, store(st, ST)\} | Trace]\})}. 

\begin{figure}[t!]
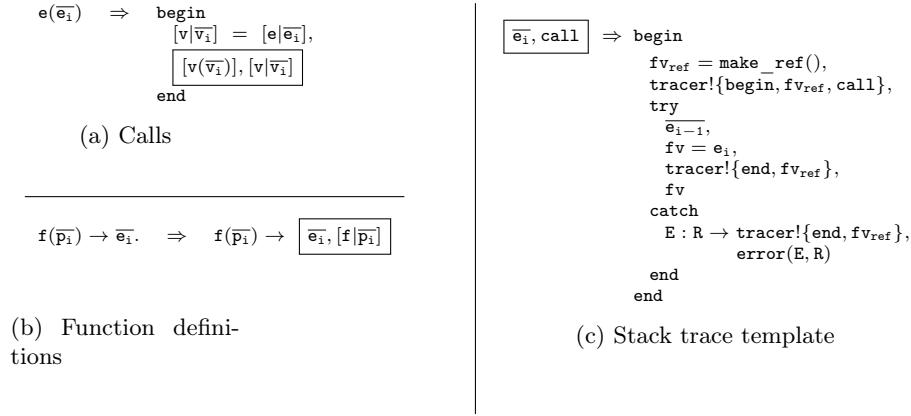

\begin{minipage}{0.5\linewidth}
\begin{subfigure}[t]{0.5\textwidth}
  \begin{minipage}{\linewidth}
  \scriptsize{
\scalebox{1.0}{
\setlength\arraycolsep{5pt} 
$\begin{array}{lll}
\mathtt{e(\overline{e_i})} & \Rightarrow & \mathtt{begin}\\
& & \hspace{6pt}\mathtt{[v | \overline{v_i}] ~=~ [e | \overline{e_i}],}\\[0.5ex]
& & \hspace{6pt}$\fbox{$\mathtt{[v(\overline{v_i})] , [v | \overline{v_i}]}$}$\\
& & \hspace{0pt}\mathtt{end}
\end{array}$
}}
  \end{minipage}
\caption{Calls}
\label{fig:stack_call}

\end{subfigure}\\[2ex]
\vspace{0.55cm}
\begin{subfigure}[t]{0.5\textwidth}
  \begin{minipage}{\linewidth}
  \scriptsize{
\scalebox{1.0}{
\setlength\arraycolsep{5pt} 
$\begin{array}{lll}
\\\hline\\
\mathtt{f(\overline{p_i})  \rightarrow \overline{e_i}.} & \Rightarrow & \mathtt{f(\overline{p_i}) \rightarrow~}$\fbox{$\mathtt{\overline{e_{i}} , [f | \overline{p_i}]}$}$
\end{array}$
}}
  \end{minipage}
\vspace{0.55cm}
\caption{Function definitions}
\label{fig:stack_fun}
\end{subfigure}%
\end{minipage}
\vline 
\begin{subfigure}[t]{0.5\textwidth}
  \begin{minipage}{\linewidth}
\vspace{-1.2cm}
  \scriptsize{
\scalebox{1.0}{
\setlength\arraycolsep{5pt} 
$\begin{array}{lll}
$\fbox{$\mathtt{\overline{e_{i}} , call}$}$ ~\Rightarrow & \hspace{-6pt}\mathtt{begin}\\
& \hspace{0pt}\mathtt{fv_{ref} = make\_ref(),}\\
& \hspace{0pt}\mathtt{tracer!\{begin, fv_{ref}, call\},}\\
& \hspace{0pt}\mathtt{try}\\
& \hspace{6pt}\mathtt{\overline{e_{i - 1}},}\\
& \hspace{6pt}\mathtt{fv = e_i,}\\ 
& \hspace{6pt}\mathtt{tracer!\{end, fv_{ref}\},}\\
& \hspace{6pt}\mathtt{fv}\\ 
& \hspace{0pt}\mathtt{catch}\\
& \hspace{6pt}\mathtt{E:R \rightarrow tracer!\{end, fv_{ref}\},}\\
& \hspace{33pt}\mathtt{error(E, R)}\\
& \hspace{0pt}\mathtt{end}\\
& \hspace{-6pt}\mathtt{end}\\
\end{array}$
}}
  \end{minipage}
\caption{Stack trace template}
\label{fig:stack_template}
\end{subfigure}
    \caption{Transformation rules to obtain the stack trace}
    \label{fig:stack_rules}
\end{figure}

\begin{wrapfigure}[13]{r}{0.45\textwidth}
\vspace{-10pt}
\begin{center}
\begin{lstlisting}[basicstyle=\ttfamily\scriptsize, frame=single, 
%caption={Comparison function which ignores the callee}, label=lst:fc_ign_callee, 
language=erlang,
numbers=left, stepnumber=1,escapechar=@]
stack_tracer(Stack) ->
  receive
    {@\color{black}{begin}@, Ref, Call} ->
      stack_tracer([{Ref, Call} | Stack]);
    {@\color{black}{end}@, Ref} ->
      case Stack of 
        [{Ref, _} | T] ->@\label{fig:stack_tracer_1_s}@
          stack_tracer(T);@\label{fig:stack_tracer_1_f}@
        [_ | T] ->@\label{fig:stack_tracer_2_s}@
          NStack = 
            unstack_till(T, Ref),
          stack_tracer(NStack)@\label{fig:stack_tracer_2_f}@
      end
  end. 
\end{lstlisting}
\vspace{-20pt}
\end{center}
\caption{Stack tracer}
\label{fig:stack_tracer}
\vspace{-10pt}
\end{wrapfigure}

However, the most challenging part here is not modifying the POI testing approach, but getting useful stack traces. Some programming language, like Erlang, perform \emph{last call optimization} (LCO) which solve important performance issues. However, LCO comes with an important drawback: the stack traces that are reported in errors can be incomplete. This can be very confusing and annoying for users when they use reported stack traces that could be used during the debugging of a buggy code. Nevertheless, there are ways to inhibit LCO. Of course, this should be done very carefully, as its impact in the performance can be disastrous. For instance, a program transformation that changes the code in a way that the last expression of function bodies is never a call. In this way, we can get full stack traces by using the standard methods provided by the language.  
An alternative is to forget these standard methods and \emph{manually} build the stack trace during the POI tracing instrumentation. 
With this approach, the stack trace is dynamically built and each time a POI trace is sent, a snapshot of the current stack (which is part of the server's state) is stored in the additional information mapping. Finally, there is another alternative which does not include the stack trace during the testing process. Instead, before UBs are reported, their correspondent ITCs are rerun sending the stack trace for only a specific execution of a POI, e.g. the fourth time it is executed.

Both alternatives, LCO inhibition and manual stack trace building, can be useful tools to get the full stack trace, especially in those cases where the expected size of the stack traces are not huge. In the case of Erlang, by inhibiting LCO we get the standard stack trace provided by Erlang, which does not include all the call arguments, but just the callee. We can improve this in the manual version. Figure \ref{fig:stack_rules} shows two transformation rules that can be used to manually obtain stack traces. This transformation should be done after the POI instrumentation in order to be correct. Both rules use a template (represented by $\fbox{$\mathtt{\overline{e_{i}} , call}$}$) to create stack traces. The idea behind this template is to send a \texttt{begin} trace just before starting the call and an \texttt{end} trace just after. This can be done in the calls (Figure \ref{fig:stack_call}) or/and in the function definitions (Figure \ref{fig:stack_fun}\footnote{The rule shows how a clause of a function definition is transformed. A whole function definition is transformed by applying this rule for each of its clauses. }). By transforming only the calls we can stack all the calls performed in user-defined code, even calls to external libraries. However, calls to user function performed from a non-user-defined code, e.g. from a \texttt{lists:map/2}, would not be stacked. By transforming only function definitions this benefit and drawback are reversed. Using both at the same time is probably the best choice. However, this configuration produces duplicated \texttt{begin-end} traces when a user-defined function is called from user-defined code. This is solved at the tracer side. Figure \ref{fig:stack_tracer} shows a simplification of a stack tracer whose state is the current stack, i.e. \texttt{Stack}. The most interesting part is how the \texttt{end} traces are processed. Clause of lines \ref{fig:stack_tracer_1_s}-\ref{fig:stack_tracer_1_f} represents the case where the top of the stack coincides with the \texttt{end} trace, i.e. a successfully finished call. On the other hand, clause of lines \ref{fig:stack_tracer_2_s}-\ref{fig:stack_tracer_2_f}, represents calls where some error has been raised during their evaluation. Function \texttt{unstack_till/2} unstack elements of the stack until the expected reference, i.e. \texttt{Ref}, is found. Errors raised during a call evaluation are the reason why the call is put inside a \texttt{try-catch} expression in the transformation rule show in Figure \ref{fig:stack_call}. The handler of this  \texttt{try-catch} expression sends an \texttt{end} trace to inform the tracer that some error has occurred.

\subsection{Using stack traces during POI traces comparison}
\label{sec:stack-use}

The inclusion of the stack trace in the comparison process unveils new challenges to the comparison modes identified in Section \ref{sec:gen-use}. 

\begin{itemize}

\item \textbf{NUAI mode:} Stack traces are only used when a discrepancy is found. This is the only mode that does not need to include the stack traces in the stacks, i.e. they can be calculated afterwards. 

\begin{itemize}

\item \textbf{NUAI-T mode:}  A new types of UB types can be generated related to stack traces. This involves the study of the stack trace. The most common way to identyfy UB types is to indicate whether the stack traces are equal till certain point or not (see Listing \ref{lst:gen_ub_types}). However, there are other UB types like that the stack traces are equal except some elements, e.g. when alternative or renamed functions are used during the calculation. 

\item \textbf{NUAI-R mode:} Using this mode, users can know whether the stack traces are equal or not for each UB detected. When they are not equal, some additional information about the discrepancy in the stack traces can be really helpful. However, we cannot always show the entire stack traces, as they are usually impractical. In order to make this information more user-friendly, a redesign of this output is needed, e.g. showing only the function calls that created a bifurcation in the stack traces.


\end{itemize}

\item \textbf{UAI mode:} This mode requires to have the stack trace information during the comparison process, i.e. it cannot be calculated afterwards. Stack trace discrepancies are usually an important reason for an UB. Therefore, stopping comparison process as soon as one of this discrepancies is found is a good strategy to find and fix the source of their discrepancy.

\item \textbf{AIT mode:} This mode requieres a special instrumentation, like the one which manually builds stack traces in Section \ref{sec:stack-trace}. This mode can be used even without defining any POI, since the idea is to have a TE each time we call/enter to a function, independently of where the POIs are placed. Therefore, this is a really interesting mode that only requires from users to define some input functions. This information is enough to check that the stack traces are equivalent in the provided tests. Of course, it could be also mixed with POI checking, merging, in this way, different checks in a one-step comparison process.  

\end{itemize}

\section{Related work}
\label{sec:rel}

The orchestrated survey of methodologies for automated software test case generation \cite{ABCCCGHHMOE13} identifies five techniques to automatically generate test cases. POI testing approach could be included in the class of  \emph{adaptive random technique as a variant of random testing}. Inside this class, the authors identify five approaches. POI testing's mutation approach of the test input shares some similarities with various of these approaches like selection of best candidate as next test case or exclusion. According to a survey on test amplification \cite{DVYMB17}, which identifies four categories that classify all the work done in the field, our work could be included in the category named \emph{amplification by synthesizing (new tests with respect to changes)}. Inside this category, our technique falls on the "other approaches" subcategory. 


The value spectra presented in \cite{XieN05} is a program spectra \cite{RBDL98} that shares several similarities with our call trace enhancement. In particular, value trace spectra record the sequence of the user-function executions traversed as a program executes. After the spectra recording, spectra comparison techniques are used to find value spectra differences that expose internal behavioral deviations inside the black box. However, the spectrum is generated for all the user-defined functions while in our approach users are who decide which functions should be compared. Additionally, POI testing approach allows a more flexible use of these call traces. Finally, the motivation and also some techniques of the enhanced call traces are similar to the ones of \emph{algorithmic debugging} \cite{Sha82}. In fact, this approach has been successfully applied in Erlang \cite{CabMRT14}.

The benefits of using stack traces in error reports are well-known. For instance, \cite{SchBP10} examined stack traces in bug reports and found that bugs are fixed faster when their reports contain at least one stack trace. Comparison of stack traces is also common in the literature. One example is \cite{BMRC05} where the authors propose a technique that can  identify similar bugs based on a comparison of stack traces.

Most of the efforts in regression testing research have been put in the regression testing minimization, selection, and prioritization \cite{YooH12}, although among practitioners it does not seem to be the most important issue \cite{EngR10}. In fact, in the particular case of the Erlang language, most of the works in the area are focused on this specific task \cite{taylor2012using,toth2013reduction}. We can find other works in Erlang that share similar goals but more focused on checking whether applying a refactoring rule will yield to a semantics-preserving new code \cite{jumpertz2010using}.

With respect to tracing, there are multiple approximations similar to the POI testing's. In Erlang's standard libraries, there are implemented two tracing modules. Both are able to trace the function calls and the process related events (spawn, send, receive, etc.). One of these modules is oriented to trace the processes of a single Erlang node \cite{dbgErlang}, allowing for the definition of filters to function calls, e.g., with names of the function to be traced. The second module is oriented to distributed system tracing \cite{ttbErlang} and the output trace of all the nodes can be formatted in many different ways. Cronqvist \cite{redbug} presented a tool named redbug where a call stack trace is added to the function call tracing, making possible to trace both the result and the call stack. Till \cite{erlyberly} implemented erlyberly, a debugging  tool with a Java GUI able to trace the previously defined features (calls, messages, etc.) but also giving the possibility to add breakpoints and trace other features such as exceptions thrown or incomplete calls. All these tools are accurate to trace specific features of the program, but none of them is able to trace the value of an arbitrary point of it. In our approach, we can trace both the already defined features and also a point of the program regardless of its position.

%
%
%
%

\section{Conclusions}
\label{sec:conclusions}


We have presented a common framework to enhance POI testing with the addition of new information. This new information enriches the approach, allowing users to get better UB reports and to define new UB types. These new UB types benefit some of the internal processes of the approach, e.g. the ITC generation.
These additions needs new ways to send and store this additional information and also new comparison modes. 
In this paper, two enhancements have been proposed by adding stack traces and augmenting call traces.  
Both enhancements have their particularities, but both share the same common framework presented in Section \ref{sec:general}.

This work opens a via to extensions of the POI testing approach. 
Following the same common framework described in this paper, we can easily include different additional information.
This additional data can be other functional data, i.e. similar data to richer call traces or stack traces. 
One interesting enhancement is to store a snapshot of the current environment for each POI, so more contextual information is available to find the UB source. 
We could also store the followed conditional paths, so it could be used to improve the coverage during the ITC generation. 
At the same time, we are studying the use of some special additional information that enables the \emph{mocking} of values. The idea is to rerun an ITC that leads to an UB, but when the value that uncovers the UB is found, replace it by the value computed by a correct version of the program. This will allow the technique to find further errors using the same ITC. The same idea can be applied when an internal call is a previously run ITC in order to avoid its recomputation.
Finally, we plan to define extensions of the approach that studies non-functional data, e.g. CPU or memory usage. 
After some preliminary work, we have concluded that the common framework presented in this paper represents a very natural way to operate with such kind of data.


%
%
%

\bibliography{\biblio{biblio}}
\bibliographystyle{abbrv}

\newpage
\appendix
\label{appendix}

The main goal of these appendices is to show the implementation, into the \texttt{SecEr} tool, of the approach presented in the paper. Concretely, we demonstrate how this additional information can be used to track down the source of a discrepancy in a more effective way. There are two use cases presented: 1) an unexpected behaviour that is located by using the enhanced call tracing, and 2) an unexpected behaviour that is detected with the help of stack traces.

\section{Use case 1: Align Columns}

This use case illustrates how the enhanced call tracing can be used to find out the source of a discrepancy. In concrete, we compare two versions of an Erlang program that aligns columns of a string with multiple lines. The code of both versions is shown in Listing~\ref{lst:align_source}. For the sake of ease the presentation, there is only one difference between both code versions. While \texttt{align\_columns\_ok.erl} version code is implemented with line \ref{lst:align_ok_line} of Listing \ref{lst:align_source}, \texttt{align\_columns.erl} version replace that line of code with line \ref{lst:align_line}. Both program versions are part of the benchmarks used in EDD (Erlang Declarative Debugger) \cite{CabMRT14}. 
\begin{figure*}[h!]
\begin{lstlisting}[tabsize=2,basicstyle=\ttfamily\scriptsize, frame=single, caption=Align columns program versions., label=lst:align_source, numbers=left, stepnumber=1,escapechar=@, otherkeywords={module,export,fun,\{,\},\[,\]}]
-module (align_columns_ok). / -module (align_columns).
-export([align_left/0, align_right/0, align_center/0]).

align_left()-> align_columns(left). @\label{lst:align_left}@
align_right()-> align_columns(right).
align_center()-> align_columns(centre).
align_columns(Alignment) -> @\label{lst:align_columns_1}@
    Lines =
         ["Given$a$text$file$of$many$lines$where$fields$within$a$line$",
          "are$delineated$by$a$single$'dollar'$character,$write$a$program",
          "that$aligns$each$column$of$fields"],
    Words = [ string:tokens(Line, "$") || Line <- Lines ],
    Words_length  = lists:foldl( fun max_length/2, [], Words),
    [prepare_line(Words_line, Words_length, Alignment) @\label{lst:align_lc_line}@
     || Words_line <- Words].

max_length(Words_of_a_line, Acc_maxlength) ->
    Line_lengths = [length(W) || W <- Words_of_a_line ],
    Max_nb_of_length = lists:max([length(Acc_maxlength), length(Line_lengths)]),
    Line_lengths_prepared = adjust_list(Line_lengths, Max_nb_of_length, 0),
    Acc_maxlength_prepared = adjust_list(Acc_maxlength, Max_nb_of_length, 0),
    Two_lengths =lists:zip(Line_lengths_prepared, Acc_maxlength_prepared),
    [ lists:max([A, B]) || {A, B} <- Two_lengths].

adjust_list(L, Desired_length, Elem) ->
    L++lists:duplicate(Desired_length - length(L), Elem).

prepare_line(Words_line, Words_length, Alignment) -> @\label{lst:prepare_line}@
    All_words = adjust_list(Words_line, length(Words_length), ""),
    Zipped = lists:zip(All_words, Words_length),
    [ apply(string, Alignment, [Word, Length + 1, $\s]) %align_columns_ok @\label{lst:align_ok_line}@
    [ apply(string, Alignment, [Word, Length - 1, $\s]) %align_columns @\label{lst:align_line}@
      || {Word, Length} <- Zipped ].
\end{lstlisting}
\end{figure*}
They can be found at: \url{https://github.com/tamarit/edd/tree/master/examples/align_columns}. These programs export only three functions with zero parameters. Thus, they can be considered unit cases. We could use any or all of these functions as starting point for the behaviour comparison process, but we will focus in just one of these three functions for simplicity. The \texttt{SecEr}'s configuration file (Listing~\ref{lst:align_config_file}), defines that the function selected as input function is \texttt{align_left/0} (Listing \ref{lst:align_config_file}, line \ref{lst:align_config_ubrm_def_funs}).

In order to increase the usability of the tool, a set of functions has been defined to easily define the \texttt{SecEr} configuration file, e.g. function call 
\texttt{secer_api:nuai_tr_config/2} in line \ref{lst:align_config_nuai_tr} of Listing~\ref{lst:align_config_file} is used to define 
NUAI-TR comparison mode.
There are also other functions used to define the VEF (Listing~\ref{lst:align_config_file}, line \ref{lst:align_config_vef_def}) or the UBRM (Listing~\ref{lst:align_config_file}, line \ref{lst:align_config_ubrm_def}).
\bigskip

\begin{figure*}[h!]
\begin{lstlisting}[tabsize=2,basicstyle=\ttfamily\scriptsize, frame=single, caption=Align columns configuration file., label=lst:align_config_file, numbers=left, stepnumber=1,escapechar=@]
-module (test_align).
-compile(export_all).
poi1Old() -> {'align_columns_ok.erl', @\ref{lst:align_left}@, call}.		 poi1New() -> {'align_columns.erl', @\ref{lst:align_left}@, call}. @\label{lst:align_config_pois_1}@ 	
poi2Old() -> {'align_columns_ok.erl', @\ref{lst:align_lc_line}@, call}.		poi2New() -> {'align_columns.erl', @\ref{lst:align_lc_line}@, call}. @\label{lst:align_config_pois_2}@ 	
poi3Old() -> {'align_columns_ok.erl', @\ref{lst:align_ok_line}@, call}.		poi3New() -> {'align_columns.erl', @\ref{lst:align_line}@, call}.

rel1() -> [{poi1Old(),poi1New()}]. @\label{lst:align_config_rel1}@ 	
rel2() -> [{poi2Old(),poi2New()}]. @\label{lst:align_config_rel2}@	
rel3() -> [{poi3Old(),poi3New()}].
funs() -> "[align_left/0]".@\label{lst:align_config_ubrm_def_funs}@

config() -> secer_api:nuai_tr_config(mytecf(),ubrm()). @\label{lst:align_config_nuai_tr}@
mytecf() ->
	fun(TO,TN) -> VEF = secer_api:vef_value_only(), @\label{lst:align_config_vef_def}@	
		            case VEF(TO) == VEF(TN) of
		            	   true -> true;
			            false ->
				            case secer_api:get_te_args(TO) == secer_api:get_te_args(TN) of
					            true -> different_value_same_args;
					            false -> different_value_different_args
				            end
		            end
	end.
ubrm() -> [{different_value_same_args,[val,ca]},{different_value_different_args,[val,ca]}]. @\label{lst:align_config_ubrm_def}@
\end{lstlisting}
\end{figure*}

In order to test the behaviour preservation between both versions with \texttt{SecEr}, it is a common practice to start by selecting as POI the last expression of each input function. Therefore, we select the POIs defined in line \ref{lst:align_config_pois_1} which are paired by the relation defined in function \texttt{rel1()} in line \ref{lst:align_config_rel1} of Listing~\ref{lst:align_config_file}.
The execution of SecEr shown in Listing~\ref{lst:align_error1} reveals an UB found in the execution of \texttt{align\_left/0}.\footnote{Note that the command \texttt{secer} is defined by a set of flags: \texttt{-pois} inputs the relation between POIs of both versions, \texttt{-funs} inputs the set of input functions, \texttt{-to} inputs the timeout given to \texttt{SecEr} (in seconds), and \texttt{-config} (unused in this example) inputs the configuration mode given to the tool. When there is no configuration defined, NUAI mode is used instead.}

\begin{figure*}[h!]
\begin{lstlisting}[basicstyle=\ttfamily\scriptsize, frame=single, caption=\texttt{SecEr} reports UB from list comprehension in line \ref{lst:align_lc_line} as POI., label=lst:align_error1,language=none, numbers=left, stepnumber=1,escapechar=@]
$ ./secer -pois "test_align:rel1()" -funs "test_align:funs()" -to 5	

Function: align_left/0
----------------------------
Generated test cases: 1
Mismatching test cases: 1 (100.0%)
  Error Types:
    + different_value => 1 Errors
        Example call: align_left()
        
------ Detected Error ------
Call: align_left()
Error Type: different_value
- - - - - - - - - - - - - - 
POI: {'align_columns_ok.erl',@\ref{lst:align_left}@,call,1}
  Trace:
    [[["Given ","a          ","text ","file   ","of     ","many     ",
       "lines      ","where ","fields ","within  ","a ","line "],
      ["are   ","delineated ","by   ","a      ","single ","'dollar' ",
       "character, ","write ","a      ","program ","  ","     "],
      ["that  ","aligns     ","each ","column ","of     ","fields   ",
       "           ","      ","       ","        ","  ","     "]]]

POI: {'align_columns.erl',@\ref{lst:align_left}@,call,1}
  Trace:
    [[["Give","a        ","tex","file ","of   ","many   ","lines    ","wher",
       "field","within",[],"lin"],
      ["are ","delineate","by ","a    ","singl","'dollar","character","writ",
       "a    ","progra",[],"   "],
      ["that","aligns   ","eac","colum","of   ","fields ","         ","    ",
       "     ","      ",[],"   "]]]
----------------------------
\end{lstlisting}
\end{figure*}

The current POI is a static call, therefore the UB source cannot be at its argument. Then, we should look for the UB source inside the function \texttt{align_columns/1}  (line \ref{lst:align_columns_1}, Listing~\ref{lst:align_source}).
In this step, one of the new features presented in this work, the enhanced call tracing, can be really helpful. Therefore, in order to use this additional information, we use the configuration defined by function \texttt{config/0} (line \ref{lst:align_config_nuai_tr}, Listing~\ref{lst:align_config_file}). This function uses the enhanced call information to classify and report new types of errors. The UB is reported again for the call POI. However, as we can observe in Listing~\ref{lst:align_error2}, this result has been extended with specific call information. The reported UB is \texttt{different_value_same_args}. Thus, it indicates that both versions performed exactly the same call as it is confirmed in the additional information provided. Therefore, according to this report, the UB source must be inside the function \texttt{prepare_line/3} (line \ref{lst:prepare_line}, Listing~\ref{lst:align_source}).

\begin{figure*}[h!]
\begin{lstlisting}[basicstyle=\ttfamily\scriptsize, frame=single, caption=\texttt{SecEr} reports UB from call to \texttt{prepare_line} in line \ref{lst:align_lc_line} as POI., label=lst:align_error2,language=none, numbers=left, stepnumber=1,escapechar=@]
$ ./secer -pois "test_align:rel2()" -funs "test_align:funs()" -to 5 -config "test_align:config()"			
Function: align_left/0
----------------------------
Generated test cases: 1
Mismatching test cases: 1 (100.0%)
  Error Types:
    + different_value_same_args => 1 Errors
        Example call: align_left()
------ Detected Error ------
Call: align_left()
Error Type: different_value_same_args
- - - - - - - - - - - - - - 
POI: {'align_columns_ok.erl',@\ref{lst:align_lc_line}@,call,1}
  Trace:
    [["Given ","a          ","text ","file   ","of     ","many     ",
      "lines      ","where ","fields ","within  ","a ","line "]]
  Call POI Info:
    Callee: prepare_line
    Args: [["Given","a","text","file","of","many","lines","where","fields","within","a","line"],
           [5,10,4,6,6,8,10,5,6,7,1,4],left]
POI: {'align_columns.erl',@\ref{lst:align_lc_line}@,call,1}
  Trace:
    [["Give","a        ","tex","file ","of   ","many   ","lines    ","wher",
      "field","within",[],"lin"]]
  Call POI Info:
    Callee: prepare_line
    Args: [["Given","a","text","file","of","many","lines","where","fields","within","a","line"],
           [5,10,4,6,6,8,10,5,6,7,1,4],left]
----------------------------
\end{lstlisting}
\end{figure*}
Then, we define a new POI inside \texttt{prepare_line} implementation. Instead of selecting as POI its last expression, i.e. the list comprehension, which we already know that behaves different, we select the expression inside this list comprehension, i.e. the call to function \texttt{apply/3} (line \ref{lst:align_ok_line}/\ref{lst:align_line}, Listing~\ref{lst:align_source}). Being this expression a function call, we can reuse the previous configuration. Listing~\ref{lst:align_error3} shows the report provided by \texttt{SecEr} with this configuration. Note that the reported UB is \texttt{different_value_different_args} now. This means that there is a discrepancy in one of the arguments between versions. By looking the UB example provided by \texttt{SecEr}, we can easily find out what argument is the UB source.

\clearpage

\begin{figure*}[t!]
\begin{lstlisting}[basicstyle=\ttfamily\scriptsize, frame=single, caption=\texttt{SecEr} reports UB from call to \texttt{apply} in lines \ref{lst:align_ok_line}/\ref{lst:align_line} as POI., label=lst:align_error3,language=none, numbers=left, stepnumber=1,escapechar=@]
$ ./secer -pois "test_align:rel2()" -funs "test_align:funs()" -to 5 -config "test_align:config2()"   

Function: align_left/0
----------------------------
Generated test cases: 1
Mismatching test cases: 1 (100.0%)
  Error Types:
    + different_value_different_args => 1 Errors
        Example call: align_left()

------ Detected Error ------
Call: align_left()
Error Type: different_value_different_args
- - - - - - - - - - - - - - 
POI: {'align_columns_ok.erl',@\ref{lst:align_ok_line}@,call,1}
  Trace:
    ["Given      "]
  Call POI Info:
    Callee: apply
    Args: [string,left,["Given",11,32]]

POI: {'align_columns.erl',@\ref{lst:align_line}@,call,1}
  Trace:
    ["Given    "]
  Call POI Info:
    Callee: apply
    Args: [string,left,["Given",9,32]]
----------------------------
\end{lstlisting}
\end{figure*}

\section{Use case 2: Mergesort}

This use case demonstrates how the stack trace can help us when looking for an UB source. Suppose that we are comparing the behaviour of two different versions of an Erlang program that implement the mergesort algorithm. The code of both versions is shown in Listing~\ref{lst:merge_source}. For the sake of ease the presentation, there is just one difference between both code versions. While \texttt{merge\_ok.erl} version code is implemented with line \ref{lst:merge_ok_call_recursive} of Listing \ref{lst:merge_source}, \texttt{merge.erl} version replace this line of code with line \ref{lst:merge_call_recursive}. Both program versions are part of the benchmarks used in EDD (Erlang Declarative Debugger) \cite{CabMRT14}. They can be found at: \url{https://github.com/tamarit/edd/tree/master/examples/mergesort}. Both programs export the function \texttt{mergesortcomp/1} which has only one parameter, i.e. a list of integers. Therefore, this function becomes our input function as it is defined in line \ref{lst:merge_input_function} of Listing~\ref{lst:merge_config_file}.

\begin{figure*}[h!]
\begin{lstlisting}[tabsize=2,basicstyle=\ttfamily\scriptsize, frame=single, caption=Mergesort program versions., label=lst:merge_source, numbers=left, stepnumber=1,escapechar=@]
-module (merge_ok). / -module (merge).
-export([mergesortcomp/1]).

-spec mergesortcomp([integer()]) -> any().
mergesortcomp(List) ->
    mergesort(List, fun comp/2).

mergesort([], _Comp) -> [];
mergesort([X], _Comp) -> [X];
mergesort(L, Comp) ->
    Half = length(L) div 2,
    L1 = take(Half, L),
    L2 = last(length(L) - Half, L),
    LOrd1 = mergesort(L1, Comp),
    LOrd2 = mergesort(L2, Comp),
    merge(LOrd1, LOrd2, Comp). @\label{lst:merge_call_line}@

merge([], [], _Comp) -> [];		@\label{lst:merge_merge_line}@
merge([], S2, _Comp) -> S2;
merge(S1, [], _Comp) -> S1;
merge([H1 | T1], [H2 | T2], Comp)  -> 	@\label{lst:merge_merge_recursive}@
        case Comp(H1,H2) of 	@\label{lst:merge_case_line}@
            false -> [H2 | merge([H1 | T1], T2, Comp)]; % merge_ok.erl @\label{lst:merge_ok_call_recursive}@
            false -> [H2 | merge(T1 ++ [H1], T2, Comp)]; % merge.erl @\label{lst:merge_call_recursive}@
            true ->  [H1 | merge(T1, [H2 | T2], Comp)]	@\label{lst:merge_true_recursive_call}@
        end.

comp(X,Y) -> X < Y.

take(0,_) -> [];
take(1,[H|_])-> [H];
take(_,[])-> [];
take(N,[H|T])-> [H | take(N-1, T)]. 

last(N, List) -> lists:reverse(take(N, lists:reverse(List))).

\end{lstlisting}
\end{figure*}

In this case, we cannot define a custom TECF as we did in the previous use case. Instead, we are just adding stack trace information to the final report of the detected UBs. For this purpose, we take profit of the defined function \texttt{secer_api:nuai_r_config/1} (line \ref{lst:merge_config_nuai_r}, Listing \ref{lst:merge_config_file}), which makes \texttt{SecEr} run in NUAI-R mode.

\begin{figure*}[h!]
\begin{lstlisting}[tabsize=2,basicstyle=\ttfamily\scriptsize, frame=single, caption=Mergesort configuration file., label=lst:merge_config_file, numbers=left, stepnumber=1,escapechar=@]
-module (test_mergesort).
poi1Old() -> {'merge_ok.erl', @\ref{lst:merge_call_line}@, call, 1}. 			poi1New() -> {'merge.erl',@\ref{lst:merge_call_line}@, call, 1}.
poi2Old() -> {'merge_ok.erl', @\ref{lst:merge_case_line}@, 'case', 1}.			poi2New() -> {'merge.erl', @\ref{lst:merge_case_line}@, 'case', 1}.
	
rel1() -> [{poi1Old(),poi1New()}].	
rel2() -> [{poi2Old(),poi2New()}].	
funs() -> "[mergesortcomp/1]". @\label{lst:merge_input_function}@

config() -> secer_api:nuai_r_config([{different_value,[val,st]}]). @\label{lst:merge_config_nuai_r}@

\end{lstlisting}
\end{figure*}

The difference between these two versions in in the recursive clause of the function \texttt{merge/3}. Therefore, it makes sense to select as POI the call to this function in line~\ref{lst:merge_call_line} of the Listing~\ref{lst:merge_source}. With the command used in Listing~\ref{lst:merge_error1},
\texttt{SecEr} provides the report shown in the same Listing. This report indicates that there are some UBs. Using the UB report, we can notice that the forth evaluation of the POI differs between both versions, while their stack is the same. 
\begin{figure*}[h!]
\begin{lstlisting}[basicstyle=\ttfamily\scriptsize, frame=single, caption=\texttt{SecEr} reports UB from call to \texttt{merge} in line \ref{lst:merge_call_line} as POI., label=lst:merge_error1,language=none, numbers=left, stepnumber=1,escapechar=@]
$ ./secer -pois "test_mergesort:rel1()" -funs "test_mergesort:funs()" -to 5 
          -config "test_mergesort:config()"

Function: mergesortcomp/1
----------------------------
Generated test cases: 5692
Mismatching test cases: 3369 (59.18%)
  Error Types:
    + different_value => 3369 Errors
        Example call: mergesortcomp([0,-1,1,2,-3])

------ Detected Error ------
Call: mergesortcomp([0,-1,1,2,-3])
Error Type: different_value
- - - - - - - - - - - - - - 
POI: {'merge_ok.erl',@\ref{lst:merge_call_line}@,call,1}
  Trace:
    [[-1,0],[-3,2],[-3,1,2],[-3,-1,0,1,2]]
  Stack
    {merge_ok,mergesort,2,{line,@\ref{lst:merge_call_line}@}}

POI: {'merge.erl',@\ref{lst:merge_call_line}@,call,1}
  Trace:
    [[-1,0],[-3,2],[-3,1,2],[-3,0,-1,1,2]]
  Stack
    {merge,mergesort,2,{line,@\ref{lst:merge_call_line}@}}
----------------------------
\end{lstlisting}
\end{figure*}
Therefore, in order to know whether the error is in the arguments of the call or in the called function, we run \texttt{SecEr} with a configuration such as the one used in the previous use case. We omit the details of this step here. The UB report indicates that the problem is inside the called function.   


Then, the next step is to place a POI inside the function \texttt{merge/3} (line \ref{lst:merge_merge_line}, Listing~\ref{lst:merge_source}). We choose the clause that contains the recursive calls (line \ref{lst:merge_merge_recursive}, Listing~\ref{lst:merge_source}) because it is the most visited clause during the evaluation. In concrete, we place the POI in the case expression in line \ref{lst:merge_case_line} of the Listing \ref{lst:merge_source}. When we rerun \texttt{SecEr}, we obtain the report shown in Listing~\ref{lst:merge_error2}. This report provides some interesting information about both POIs. The behaviour discrepancy has been detected in the values computed in the fifth evaluation of the POI. Additionally, both stack traces differ. In the stack trace produced by the old version there are two stacked calls to function \texttt{merge/2} 
while in the stack trace of the new one there is only one. This means that the old version is performing an extra recursive call before reaching the base case. Then, the final step is to place a POI in each recursive call observing the values of their arguments as in the previous case. With the report provided by \texttt{SecEr} when using this configuration, users can easily spot the UB source.

\begin{figure*}[h!]
\begin{lstlisting}[basicstyle=\ttfamily\scriptsize, frame=single, caption=\texttt{SecEr} reports UB from case expression in line \ref{lst:merge_case_line} as POI., label=lst:merge_error2, language=none, numbers=left, stepnumber=1, escapechar=@]
$ ./secer -pois "test_mergesort:rel2()" -funs "test_mergesort:funs()" -to 5 
          -config "test_mergesort:config()"

Function: mergesortcomp/1
----------------------------
Generated test cases: 4878
Mismatching test cases: 2885 (59.14%)
  Error Types:
    + different_value => 2885 Errors
        Example call: mergesortcomp([5,-6,-6,2,3])

------ Detected Error ------
Call: mergesortcomp([5,-6,-6,2,3])
Error Type: different_value
- - - - - - - - - - - - - - 
POI: {'tests/mergesort/merge_ok.erl',@\ref{lst:merge_case_line}@,'case',1}
  Trace:
    [[-6,5],[2,3],[-6,2,3],[3,5],[2,3,5]]
  Stack
    {merge_ok,merge,3,{line,@\ref{lst:merge_true_recursive_call}@}}
    {merge_ok,merge,3,{line,@\ref{lst:merge_ok_call_recursive}@}}
    
POI: {'tests/mergesort/merge.erl',@\ref{lst:merge_case_line}@,'case',1}
  Trace:
    [[-6,5],[2,3],[-6,2,3],[3,5],[-6,3,5]]
  Stack
    {merge,merge,3,{line,@\ref{lst:merge_call_recursive}@}}
----------------------------

\end{lstlisting}
\end{figure*}

\end{document}

\newpage
\noindent \underline{Note for the reviewers:} The following appendix has been only included to ease the reviewing process, and it will not be part of the final paper. In case of acceptance, this appendix will be published as a technical report so that the interested reader will have public access to it.

\input{./Secciones/Pruebas.tex}

\end{document}